\documentclass[reprint,
 amsmath,amssymb,
 aps,
pre,
floatfix,
]{revtex4-2}

\usepackage{graphicx}
\usepackage{dcolumn}
\usepackage{bm}
\usepackage{color}
\begin{document}

\title{Application of the Variational Autoencoder to Detect the Critical Points of the Anisotropic Ising Model}

\author{Anshumitra Baul}
\affiliation{Department of Physics and Astronomy, Louisiana State University, Baton Rouge, Louisiana 70803, USA}

\author{Nicholas Walker}
\affiliation{Lawrence Berkeley National Laboratory, 1 Cyclotron Rd, Berkeley, California 94720, USA}

\author{Juana Moreno}
\affiliation{Department of Physics and Astronomy, Louisiana State University, Baton Rouge, Louisiana 70803, USA}%
\affiliation{Center for Computation and Technology, Louisiana State University, Baton Rouge, LA 70803, USA}

\author{Ka-Ming Tam}
\affiliation{Department of Physics and Astronomy, Louisiana State University, Baton Rouge, Louisiana 70803, USA}%
\affiliation{Center for Computation and Technology, Louisiana State University, Baton Rouge, LA 70803, USA}

\date{\today}

\begin{abstract}
We generalize the previous study on the application of variational autoencoders to the two-dimensional Ising model to a system with anisotropy. Due to the self-duality property of the system, the critical points can be located exactly for the entire range of anisotropic coupling. This presents an excellent test bed for the validity of using a variational autoencoder to characterize an anisotropic classical model. We reproduce the phase diagram for a wide range of anisotropic couplings and temperatures via a variational autoencoder without the explicit construction of an order parameter. Considering that the partition function of $(d+1)$-dimensional anisotropic models can be mapped to that of the $d$-dimensional quantum spin models, the present study provides numerical evidence that a variational autoencoder can be applied to analyze quantum systems via Quantum Monte Carlo.

\end{abstract}

\maketitle

\section{Introduction}
Machine learning (ML) has become an indispensable tool to reach the boundaries of scientific understanding in the age of big data. An overflow of information is being analyzed using ML to quantify patterns in a large variety of fields from  social networking, object and image recognition, advertising, finance, engineering, medicine, biological physics, and astrophysics among others~\cite{Alom2018TheHB}.

Machine learning is a data modeling approach which employs algorithms which favor strategies  driven by statistical analysis and based on pattern extraction and inference. ML algorithms, such as deep learning, provide new advances to understand physical data. Opportunities for scientific investigations are being devised particularly in numerical studies, which naturally involve large data sets and complex systems, where explicit algorithms are often challenging. A concerted effort to address large data problems in statistical mechanics and many-body physics using the ML approach is emerging \cite{Carrasquilla_2017,struc_clss,melting,ml_pt,ml_pt_lat,wetzel2017machine,PhysRevB.95.035105, Carrasquilla_2017,wetzel2017machine,torlai2016learning}. The foundation of ML is deeply connected with statistical physics, and hence is fruitful to combine ML techniques with numerical methods which involve the prediction of phase transition regions.  Scaling and renormalization are the core principles to understand macroscopic phenomena from microscopic properties. The way forward for machines to learn from large data sets would incorporate conceptually similar principles \cite{torlai2016learning,ml_rg}.

Changes in the macroscopic properties of a physical system occur at phase transitions, which often involve a symmetry breaking process \cite{Landau_1937}. The theory of phase transitions and symmetry breaking was formulated by Landau as a phenomenological model, and was later devised from microscopic principles using the renormalization group \cite{Fisher_1974}. Phases can be identified by an order parameter, which is zero in the disordered phase and finite in the ordered phase. The order parameter is determined by symmetry considerations of the underlying Hamiltonian. There are states of matter where the order parameter can only be defined in a complicated non-local way. These systems include topological states such as quantum spin liquids ~\cite{Savary_Balents_2016}. A major goal of the ML approach 
in complex statistical mechanics models or in strongly correlated systems is to detect the phase transitions from the data itself without explicitly constructing any order parameter \cite{Carrasquilla_2017}. 

The development artificial neural networks to detect phase transitions is a major advance in the area of ML applications in physics. In earlier works, artificial neural networks were based on supervised learning algorithms \cite{Carrasquilla_2017,torlai2016learning}. 
Labeled data is used to train the supervised learning algorithm, from which the algorithm learns to assign labels to the input data points ~\cite{schmidhuber2015deep,ayodele2010types}. Apart from supervised learning, another major type of ML is unsupervised learning for which the input data has no label.
Conventional unsupervised learning algorithms, such as principal component analysis \cite{ml_pearson}, find structure in unlabeled data without involving any artificial neural network. Here, the data is classified into clusters and labels can then be assigned to the data points accordingly \cite{ml_pearson}. 

Autoencoder is a new direction to utilize artificial neural networks in unsupervised machine learning. The very first versions of the autoencoder were being used for dimensional reduction of data before feeding its output into other ML algorithms~\cite{wang2016auto,wang2012folded}. It is created by encoding an artificial neural network, which outputs a latent representation of the given data, and is utilized as the input of decoding neural network that tries to accurately reconstruct the input data from the latent representation ~\cite{almotiri2017comparison,plaut2018principal}.

A major shortcoming of autoencoder is the possibility of sharp changes in the latent representation with respect to small differences in the input data. Ideally, the latent representation should be a smooth function of the input data. The variational autoencoder (VAE) represents the latent representation in terms of probability distributions instead of a fixed set of numbers \cite{Kingma2014AutoEncodingVB,kingma2019introduction}. This probabilistic latent representation allows a smooth latent representation. Since 2013, VAEs have been developed to become one of the most successful unsupervised learning algorithms \cite{Kingma2014AutoEncodingVB}. Promising results are being shown in both encoding and reconstructing data \cite{ising_vae,Kingma2014AutoEncodingVB,kingma2019introduction}.

VAEs are being successfully applied recently to detect phase transitions in classical spin models \cite{ising_vae,ising_vae_2,ising_vae_3}. The input data sets are given by Monte Carlo method and then unsupervised machine learning, such as the VAE is used for deciphering and distinguishing different physics in the input data sets. After successfully applying to classical models, a natural question arises, whether such an approach remains viable for quantum models. In particular, can a VAE be viable to distinguish different quantum phases and transition regions based on the obtained data from quantum Monte Carlo. Recently, various models from statistical mechanics, in particular the Ising model and the Potts model have been investigated~\cite{kim2018smallest,morningstar2017deep,Shiina_etal_2020}. In this work, we investigate a rather simple quantum model, one-dimensional transverse field Ising model(TFIM), to address the capability and the limitations of the autoencoder. Given that, the critical line of the model can be calculated analytically due to self-dual property ~\cite{kwd_2,kwd_3,kwd}. The TFIM is an excellent test bed to address the various aspects of VAEs in quantum models.

This paper is organized as follows. In section II, we briefly describe the transverse field Ising model (TFIM) and the Suzuki-Trotter formulation by mapping it to the anisotropic Ising model. In section III, the Monte Carlo method and the VAE are presented. The results from the VAE are described in the section IV. We conclude and discuss the implication and possible future applications of the method developed in this study in Section V. The self-duality of the anisotropic two-dimensional Ising model and the detail of the VAE are discussed in the appendix.

\section{Transverse field Ising Model}
\subsection{Model}
We consider an Ising model with transverse field ~\cite{PFEUTY197079,Stinchcombe_1973,p_cond_mat}. 
The Hamiltonian is given as

\begin{equation}
    H=-\sum_{<i,j>}J_{ij}\sigma^{z}_{i}\sigma^{z}_{j}-\Gamma\sum_{i}\sigma^{x}_{i},
\end{equation}
where $\sigma^{\alpha}(\alpha=x,y,z)$ are the Pauli matrices, which obey the commutation relation, $[\sigma^{\alpha}_{i},\sigma^{\beta}_{j}]=2 \iota \delta_{ij} \epsilon_{\alpha \beta \gamma}\sigma^{\gamma}_{i}$. $J_{ij}$ is the coupling between the spins at sites $i$ and $j$. Only nearest neighbors coupling is considered in this study. $\Gamma$ is the transverse field applied in the $x$-direction. $\iota$ is an imaginary number, and $\sigma^{z}$ has the eigenvalues as $\pm 1$. Their eigenvectors are symbolically denoted by $|\uparrow>$. and $|\downarrow>$, that is 
\begin{equation}
    |\uparrow> = \begin{pmatrix} 1 \\
    0
    \end{pmatrix}
\end{equation}
and 
\begin{equation}
    |\downarrow> = \begin{pmatrix} 0 \\ 
    1
    \end{pmatrix}.
\end{equation}
The order parameter is given by the average magnetization $m= \sum_{i}<\sigma^{z}_{i}>/N$ ($N$ being the total number of sites) which characterizes the phase transition between paramagnetic and ferromagnetic phases. Without specification, we consider only the one dimensional TFIM with coupling limited to the nearest neighbors. 

\subsection{Suzuki-Trotter Formalism}
Instead of working with the quantum spins directly, we follow the standard procedure of mapping a $d$-dimensional quantum Hamiltonian into a $d+1$-dimensional effective classical Hamiltonian by the Suzuki-Trotter transformation \cite{suzuki2012quantum,suzuki1971relationship,chakrabarti2005transverse}. We apply this to the TFIM system. We first define the longitudinal spin coupling terms and the transverse filed terms as follow  
\begin{eqnarray}
    H_{0}&\equiv&-\sum_{<i,j>}J_{ij}\sigma^{z}_{i}\sigma^{z}_{j}. \\ \nonumber
    V&\equiv&=-\Gamma\sum_{i=1}\sigma^{x}_{i}. \\ \nonumber
    H&=&V+H_{0}.
\end{eqnarray}
The partition function of $H$ reads 
\begin{eqnarray}
 Z=Tr(e^{-\beta(H_{0}+V)}),
\end{eqnarray}
$\beta$ is the inverse temperature.
The Trotter formula gives,
\begin{equation}
    exp(A_{1}+A_{2})=\lim_{M \rightarrow \infty}[exp(A_{1}/M)exp(A_{2}/M)]^{M}
\end{equation}
when $[A_{1},A_{2}] \neq 0$.
Using the trotter formula we have
\begin{equation}
    Z=\sum_{i}\lim_{M\rightarrow \infty}\langle s_{i}|[exp(-\beta H_{0}/M)exp(-\beta V/M)]^{M}|s_{i} \rangle,
\end{equation}
where $s_{i}$ is the $i$-{th} spin configuration of the whole system, and the above summation runs over all the $2^N$ possible configurations denoted by $i$. $M$ identifies the number of identity operators. The identity operator formed from the spin operators is given as,
\begin{eqnarray}
 I &&=\sum^{2^N}_{i}|s_{i,k} \rangle \langle s_{i,k}| \\ \nonumber
   &&=\sum^{1}_{\sigma_{i,k}=-1}| \sigma_{i,k},...,\sigma_{N,k} \rangle \langle \sigma_{i,k},...,\sigma_{N,k}| 
\end{eqnarray}
where $k=1,2,...,M$. Hence, it is the product of $M$ exponentials in $Z$.
\begin{eqnarray}
    Z=&&\lim_{M\rightarrow \infty}Tr \prod^{M}_{k=1} \langle \sigma_{i,k},...,\sigma_{N,k}| \\ \nonumber
   &&exp(-\frac{\beta H_{0}}{M}) exp(-\frac{\beta V}{M})|\sigma_{i,k+1},...\sigma_{N,k+1} \rangle \\ 
\end{eqnarray}
Applying the periodic boundary conditions implies
$\sigma_{N+1,p}=\sigma_{1,p}$.
After evaluating the expression of the partition function,
\begin{equation}
    Z=C^{\frac{NM}{2}}Tr_{\sigma}(-\beta H_{eff}[\{\sigma]\}),
\end{equation}
where $C=\frac{1}{2}\sinh{\frac{2\beta \Gamma}{M}}$ and the effective classical Hamiltonian
\begin{eqnarray}
    H_{eff}(\{\sigma\})=&&\sum^{N}_{<i,j>}\sum^{M}_{k=1}[\frac{-J_{ij}}{M}\sigma_{i,k}\sigma_{j,k}  \\ \nonumber&&-\frac{\delta_{ij}}{(2\beta)} \ln {(\coth{\frac{\beta \Gamma}{M}})} \sigma_{i,k}\sigma_{i,k+1}].
\end{eqnarray}
$\sigma_{i,k}$ involved are the eigenvalues of $\sigma^{z}$ and hence there is no non-commuting part in $H_{eff}$. The effective Hamiltonian shows the system of spins in the $d+1$-dimensional lattice and one extra label $k$ for each spin variable. Each single quantum spin variable $\sigma_{i}$ in the original Hamiltonian, now we have an array of $M$ number of classical spins $\sigma_{i,k}$. Hence, the partition function of the 1D quantum Ising model is mapped to that of the 2D Ising model. This new (time-like) dimension along which these classical spins are spaced is called the Trotter dimension. 

In this paper we assume the model only has nearest neighbor coupling on a square lattice. The coupling along the $x$-direction and $y$-direction are denoted as $J_{x}$ and $J_{y}$ respectively. We also define $K_{x}=1/(\beta J_{x})$ and $K_{y}=1/(\beta J_{y})$. We set $N=M$ in this study.

For the two-dimensional classical Ising model, the critical points can be obtained due to the self-dual property. The detail of the dual transformation is shown in the appendix A. 

\section{Methodology}

\subsection{Monte Carlo Sampling}
 The spin configurations are generated by using the single spin-flip Metropolis algorithm. In a single spin flip, first we flip the spin of a single lattice site, and then calculate the change in energy $\Delta E$. After that, we use the resulting change in energy to calculate the Metropolis criterion $exp(-\Delta E/T)$.  If a randomly generated uniform number is in the range of $[0,1)$, and is smaller or equal to the Metropolis criterion, the configuration is accepted as the new configuration. The code for the simulation is written in Python using the NumPy library \cite{python,numpy,dask,numba}.
 We note that using generative neural network, instead of Monte Carlo method, for sampling has also been proposed recently for the 1D quantum spin models \cite{Nicoli_etal_2020,Vielhaben_Nils_2021}.
 
\subsection{Variational Autoencoder}

Variational autoencoder(VAE) is under the category of generative models. New data can be produced by learning the distribution of the input data ~\cite{doersch2021tutorial,Kingma2014AutoEncodingVB}. We use an encoder-decoder architecture where the encoder maps the input to a latent representation in term of some chosen distributions. The latent distribution is mapped back to reconstruct the input using the decoder. The latent representation in a well-trained model can be used to generate new samples resembling the original training data ~\cite{larsen2016autoencoding}. 

In this study, we only consider a multi-dimensional Gaussian distribution for the latent space representation. We briefly explain the main concepts of the VAE in the following. 

\textbf{Encoder}: Encoder is a neural network that takes the higher dimensional input data into a lower dimensional latent space. For example, an image of the spin configuration of size $32\times32 = 1024$ lattice points, can be converted into a vector of dimension $8$. The encoder is a method for dimensionality reduction. The neural network representing the encoder maps each sample to a distribution and is also called a probabilistic encoder of VAE. The encoder is denoted as $q_\phi(z|x)$, which is a distribution that maps an input sample $x$ (the Ising spin configurations) to provide a latent representation $z$ (the parameters in the multi-dimensional Gaussian distribution). $\phi$ is a set of learnable parameters in the neural network that is varied to produce outputs from the encoder. 

\textbf{Latent Space}: The latent space is the input for the decoder network, and it is outputted from the encoder network. For the VAE, the latent space is represented by multi-dimensional Gaussian distribution. As the Gaussian distribution is completely specified by its mean and standard deviations, the dimension of the latent space is two times as that of the dimension of the multi-dimensional Gaussian distribution. An encoded sample is denoted by $z$.

The latent space is regularized and then get penalized for deviating from the prior multi-dimensional Gaussian distribution by the Kullback-Leibler (KL) divergence term \cite{kullback1951information}. 

\begin{figure}[tbh]
   \centering
   \includegraphics[height=50mm]{{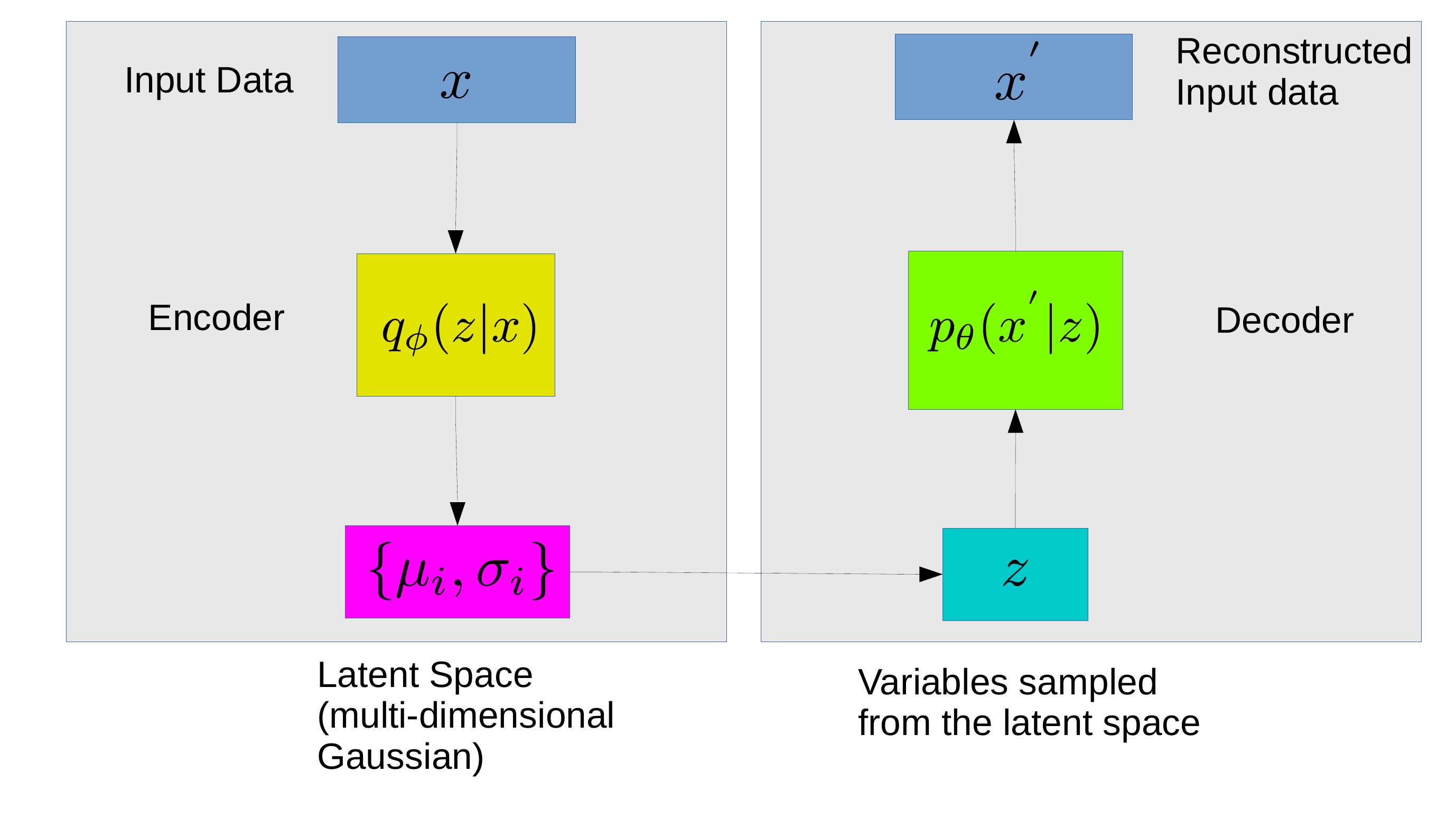}}
  \caption{Diagram depicting the structure of the VAE. The left hand side is the encoding part and the  right hand side is the decoding part. $x$ is the input Ising spin configuration, $q_\phi(z|x)$ is the encoder neural network, $\mu$ and $\sigma$ are the latent means and standard deviations of the latent space distribution, $z$. $p_\theta(x^{'}|z)$ is the decoder neural network and $x'$ is the reconstructed Ising spin configuration.}
  \label{fig:1}
\end{figure}

\textbf{Decoder}: From the fig. \ref{fig:1} one can see that the decoder in VAE converts the compressed samples in the latent space back to the original input samples~\cite{weng2018VAE} . It is represented as $p_\theta(x|z)$, a distribution to produce reconstructed samples $x'$ conditioned on latent representations $z$. $\theta$ is a learnable parameter in the neural network that can be varied to produce different outputs from the decoder. A sampling distribution provides the input to the decoder which is described by the latent representation obtained from the encoder.

\textbf{Loss Functions in VAE}: 
The VAE contained two sets of trainable parameters, $\theta$ and $\phi$ for the neural networks of the encoder and decoder respectively. They are trained by minimizing the loss function. The loss functions in VAE consist of two terms. The first one measures the ``similarity" of the inputs and the reconstructed outputs. The second one measures the difference between the designated priori distribution, chosen to be multi-dimensional Gaussian distribution, and the actual distribution of the compressed inputs.

For the first term of the loss function, the standard reconstruction loss measures the error between the samples generated by the decoder and the original input samples. It is measured by the binary crossentropy between the encoder input and decoder output in our study. This is expressed as $L_{RC}=-E_{z \sim q_\phi(z|x)}[ \log p_\theta(x|z) ]$ The expectation, $E$, is over the representations, $z$ with respect to the encoder's distribution. 

The second is the Kullback-Leibler divergence (KLD) of the latent representations. KLD measures the divergence between the chosen latent representation $p(z)$ and the approximated distribution from the output of the encoder $q_\phi(z|x)$. 

The KLD is defined as
\begin{equation}
    L_{KLD}=D_{KL}\left[ q_\phi(z|x)||p(z)\right]=-\sum_z q_\phi(z|x)\log\left(\frac{q_\phi(z|x)}{p(z)}\right).
\end{equation}
It is minimized to optimize the latent representation from the encoder $q_\phi(z|x)$ to resemble the latent representations from $p(z)$.

The total loss is the sum of the reconstruction loss and the KLD ~\cite{burgess2018understanding},
\begin{eqnarray}
    L(\phi,\theta ;x,z)=-E_{z \sim q_\phi(z|x)}\left[ \log p_\theta(x|z) \right] \nonumber \\
    +D_{KL}\left[q_\phi(z|x)||p(z)\right]
    =L_{RC}+L_{KLD}.
\end{eqnarray}
The two losses in the VAE together are optimized simultaneously to describe the latent state for an observation with distributions close to the prior approximated distribution but also deviates to describe the salient features of the input.

The linear combination of the reconstruction loss and the KLD is often denoted as a variational lower bound or evidence lower bound loss function as both the reconstruction loss and the KLD are non-negative. Minimizing the loss, $min_{\theta,\phi}L(\theta,\phi;x,z)$, maximizes the lower bound of the probability of generating new samples ~\cite{larsen2016autoencoding}.

\subsection{$\beta$-total correlation VAE}

A further refinement of the VAE can be achieved by decomposing the KLD into three parts. The three parts describing the index-code mutual information, total correlation, and dimension-wise Kulback-Leibler divergence, which are characterized by parameters often denoted as $\alpha$, $\beta$, and $\gamma$ respectively. $\beta$ is the most important one for obtaining good results \cite{chen2019isolating,burgess2018understanding}. 

A $\beta$- total correlation VAE ($\beta$-TCVAE) is defined as the VAE with the $\alpha=\gamma=1$ and tuning $\beta$ as a parameter. The detail is given in the appendix B. It is well suited for representation learning of patterns \cite{chen2019isolating}. We fix the parameters of the decomposition with $\alpha=\gamma=1$ and $\beta=8$ in this study, these values were found to be good for finding the phase diagram of the 2D isotropic Ising model \cite{walker2020identifying,ising_vae_3}.

Our goal is to map the raw Ising spin configurations to a reduced set of descriptors that discriminate between the samples using the criterion inferred by the $\beta$-total correlation VAE \cite{walker2020identifying,ising_vae_3}. The encoder and decoder are implemented as deep convolution neural networks (CNN) to preserve the spatially dependent $2$-dimensional structure of the Ising spin configurations \cite{cnn}. Scaled exponential linear unit (SELU) activation function is used in each convolution layer. The output of the final convolution layer is flattened, and fed into two $2$-dimensional dense layers. Then, it is used as the input layer for the decoder CNN and reshaped to match the structure of the output from the final convolution layer in the encoder CNN. Hence, the decoder network is simply the opposite of the encoder network, incorporated with convolution transpose layers in favor of standard convolution layers ~\cite{cnn,snn,walker2020identifying}. The final output layer from the decoder network is reproduced from the original input configurations obtained from the encoder network which uses a sigmoid activation function. The loss term consists of the reconstruction loss and the $\beta$-total correlation KLD ($\beta$-TCKLD) term with $\alpha=\gamma=1$ and $\beta=8$ ~\cite{kullback1951information,walker2020identifying}. We employ minibatch stratified sampling on the given data while training. 

To optimize the loss, Nesterov-accelerated Adaptive Moment Estimation(Nadam) was used and it efficiently minimizes the loss during the training of the $\beta$-TCVAE model ~\cite{sgd,walker2020identifying}. The default parameters provided by the Keras library and a learning rate of $0.00001$ were chosen. The training is performed over $100$ epochs with a batch size of $33 \times 33 = 1089$ for both lattice size $N=64$ and $N=128$ with a number of sample $1024$ per phase point. The reduced descriptors of the 2D Ising spin configuration are given by the latent variables ~\cite{walker2020identifying}.
The $\beta$-TCVAE model used in this work was implemented using the Keras ML library with TensorFlow as a supporting backhand \cite{tf,keras}.

\subsection{Principal Component Analysis on the Latent Space} 

The principal components analysis (PCA) is applied on the latent means and standard deviations of the Ising spin configurations obtained after fitting the $\beta$ -TCVAE, independently to produce linear transformations of the Gaussian parameters that discriminate between the samples using the scikit-learn package ~\cite{ml_pearson,van2008visualizing}. The PCA performs an orthogonal transformation into a new basis set of linearly uncorrelated features, principal components. Each principal component encompasses the largest possible variance across the sample space under an orthogonality constraint ~\cite{ml_pearson}.

The latent representation characterizes the structure of the Ising configurations, but the principal components of the latent representation shows greater discrimination between the different structural characteristics of the configurations compared to the raw variable representations ~\cite{walker2020identifying}. The rationale of using the PCA is to provide a more compact representation which characterizes the different phases of the Ising model. As we will show in the results section, the first and the second dominated components already show good signals to distinguish different phases of the anisotropic Ising model. 

\section{Results}
As explained in the previous section, for the VAE models, the samples were drawn from a multi-dimensional Gaussian distribution parametrized by vectors of means and standard deviations denoted as $\mu_i$ and $\sigma_i$ respectively, where $i$ is the index of the dimension of the distribution. All of the plots in this paper were generated with the MatPlotLib package using a perceptually uniform colormap \cite{matplotlib}. In each plot, the coloration of each square pixel on the diagram represents the ensemble average value of the principal components of the latent space mean or variance of the multi-dimensional Gaussian function. We study two different system sizes $N=64$, and $128$. We will focus on the first two principal components of the latent variance and the second principal component of the latent mean. We denote them as $\tau_0$, $\tau_1$, and $\nu_1$ respectively. The first principal component of the latent mean, $\nu_0$ does not capture a clear distinction between the ferromagnetic and paramagnetic phases of the system.

$\nu_1$ are plotted in fig. \ref{fig:2}. It is apparent that $\nu_1$ is a resemblance of the magnetization, $m$, of snapshots of the Ising spin configuration. We note that it is not to be expected that any of the latent variables to have the same value of any physical quantities, such as magnetization. Nonetheless, from the plot of $\nu_1$, it clearly discriminates  the ferromagnetic from paramagnetic phases, separating them by the phase transition line. The phase transition line in white corresponds to the analytical solution, eq. \ref{eq:exact} in the appendix calculated in the thermodynamic limit. Since the magnetization can be seen as the order parameter for the $2$-dimensional Ising model, a reasonable representation of the order parameter is seen to be possibly extracted from the VAE. We note that the simulations and the VAE are done on finite size systems, a true broken symmetry does not occur, that is the reason of the seemingly random values of $\nu_1$ in the ferromagnetic phase. Usually, the amplitude of the magnetization is considered for finite size simulation, as we will show other latent variables from the VAE has similar structure as the amplitude of the magnetization. As the magnetization is a linear feature of the Ising spin configuration, a simpler linear model would be sufficient for extracting the magnetization. 

The first principal component of the latent variance $\tau_0$ are plotted in fig \ref{fig:4}. The white line is the analytical solution for the phase transition points. One can find that the value of $\tau_0$ remains very small in the upper right region of the figures. This can be considered as the reflection of the small changes of amplitude of the energy or amplitude of the magnetization in the paramagnetic phase. Once the systems approach the critical line from the upper right region of the figure, the value of $\tau_{0}$ increase sharply. This behavior is again consistent with the amplitude of energy or magnetization. 

In particular, we consider the case for the isotropic limit ($K_x = K_y$), that is the classical ($\Gamma=0$) limit. The critical point is given as $K_c(=K_x=K_y) = \frac{2}{\ln[1+\sqrt{2}]}  \sim 2.2721$ ~\cite{onsager}. From the fig. \ref{fig:4}, we see a sharp change in the $\tau_0$ around the value of $K_c$. This result is consistent with the accomplishments of prior published works for the isotropic Ising model on a square lattice \cite{ising_vae,ising_vae_2,ising_vae_3,walker2020identifying}. 

\begin{center}

\begin{figure*}[tbh]
    \centering
        \includegraphics[width=8.525cm]{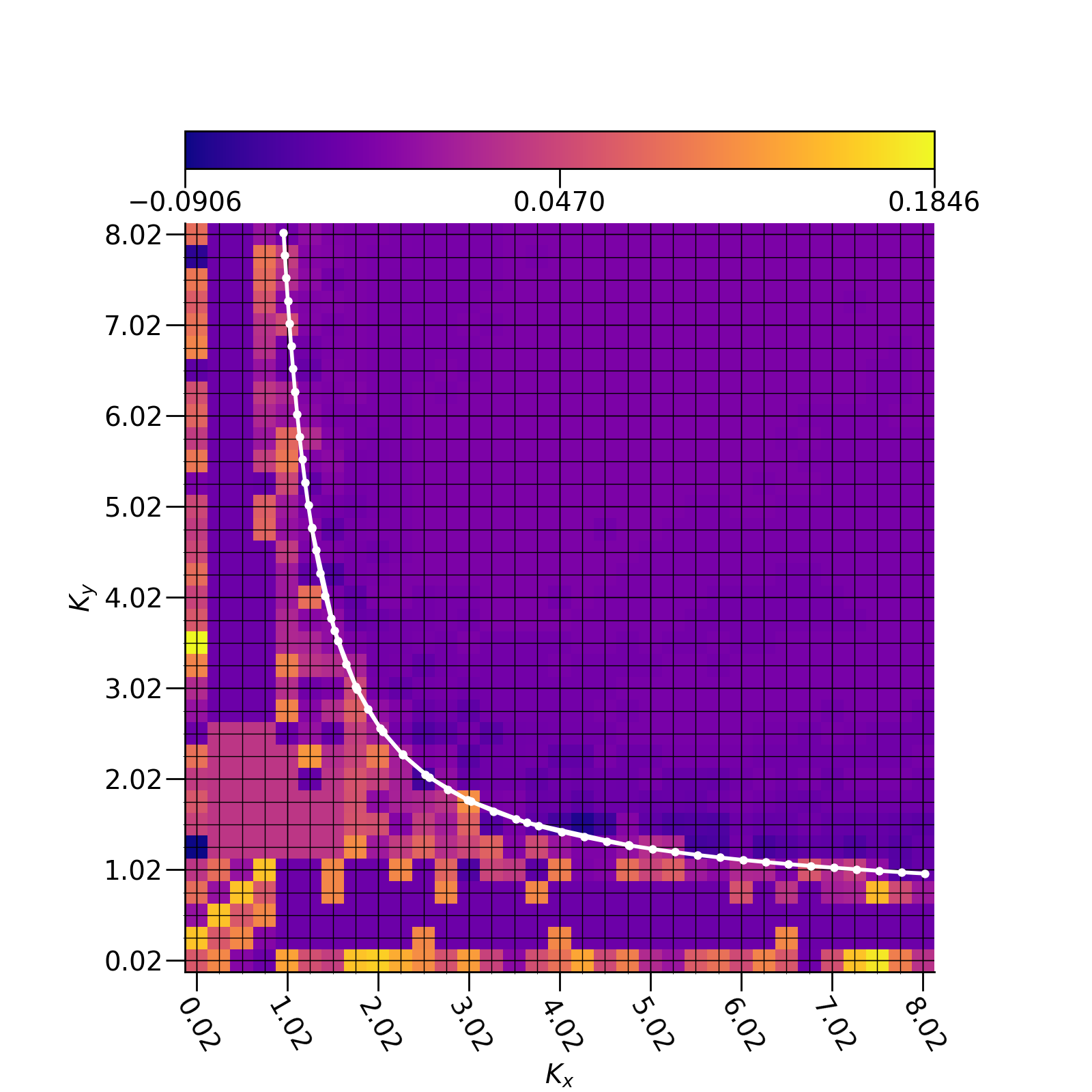} 
        \qquad
        \includegraphics[width=8.525cm]{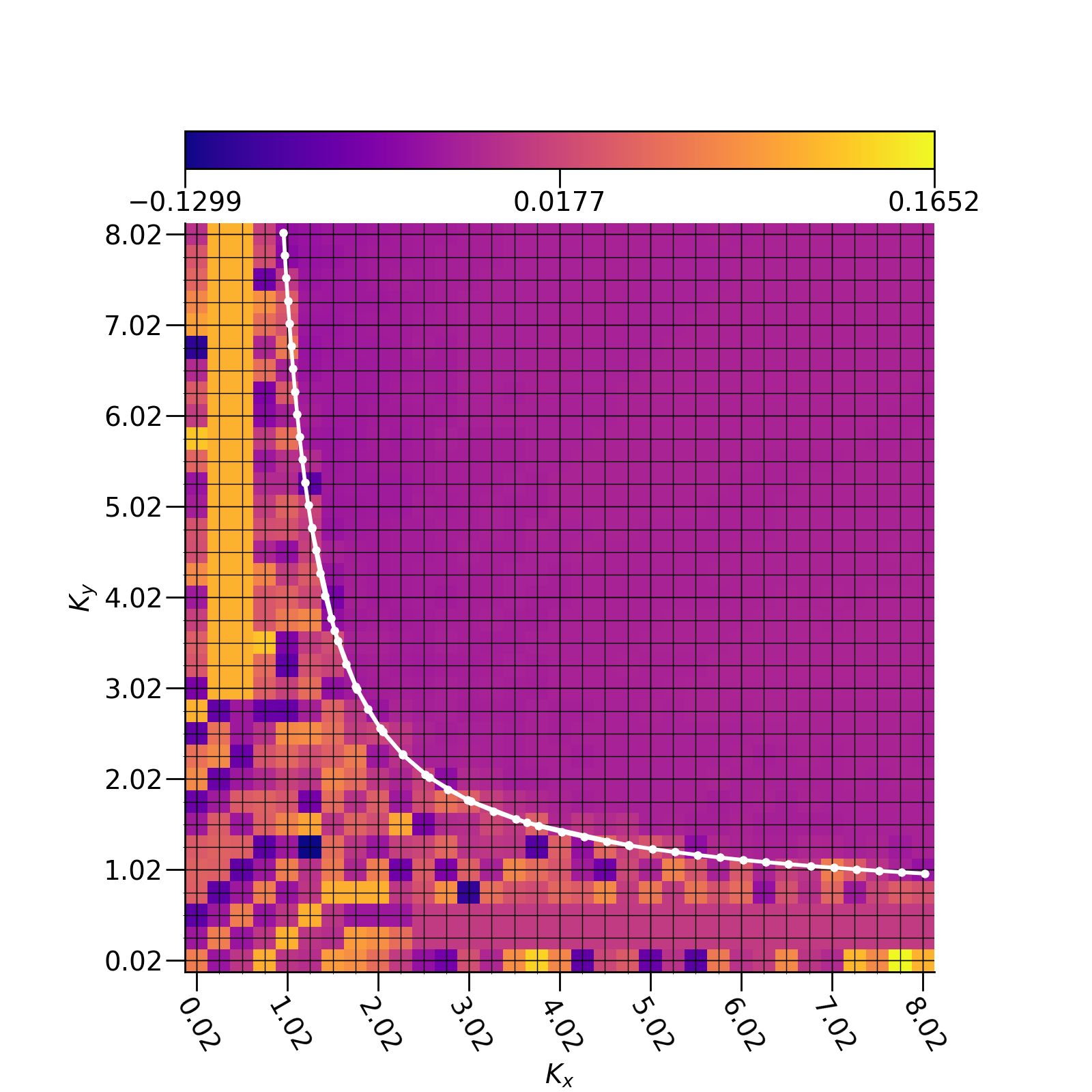} 
    \caption{The second principal component of the latent mean, $\nu_1$, with respect to the different $K_x$ and $K_y$ ($1/(\beta J_x)$ and $1/(\beta J_y)$) for the $2D$ square lattice. The left panel and right panel are for the system sizes $N=M=64$ and $N=M=128$, respectively.}
    \label{fig:2}
\end{figure*}

\begin{figure*}
    \centering
     \includegraphics[width=8.525cm]{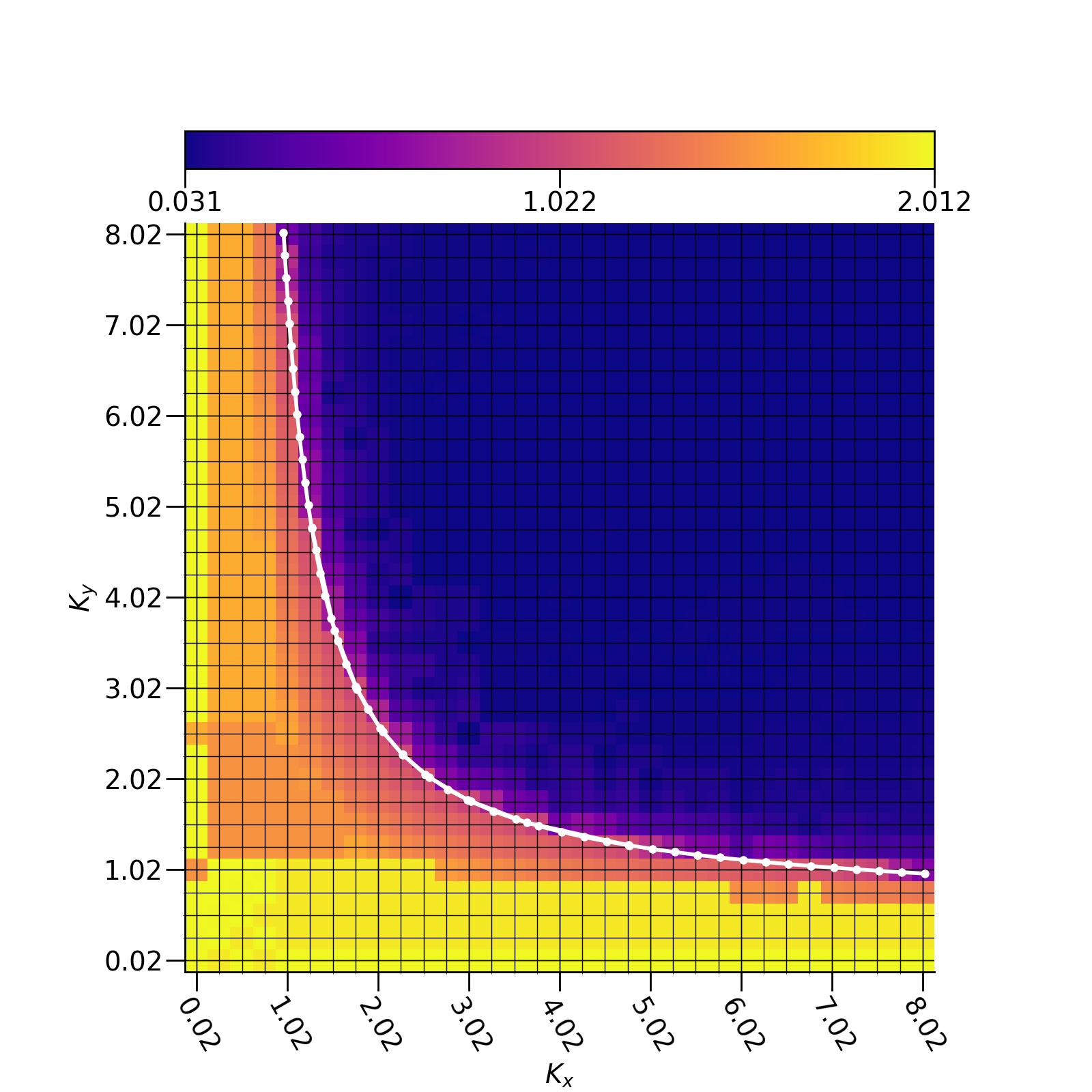} 
     \qquad
     \includegraphics[width=8.525cm]{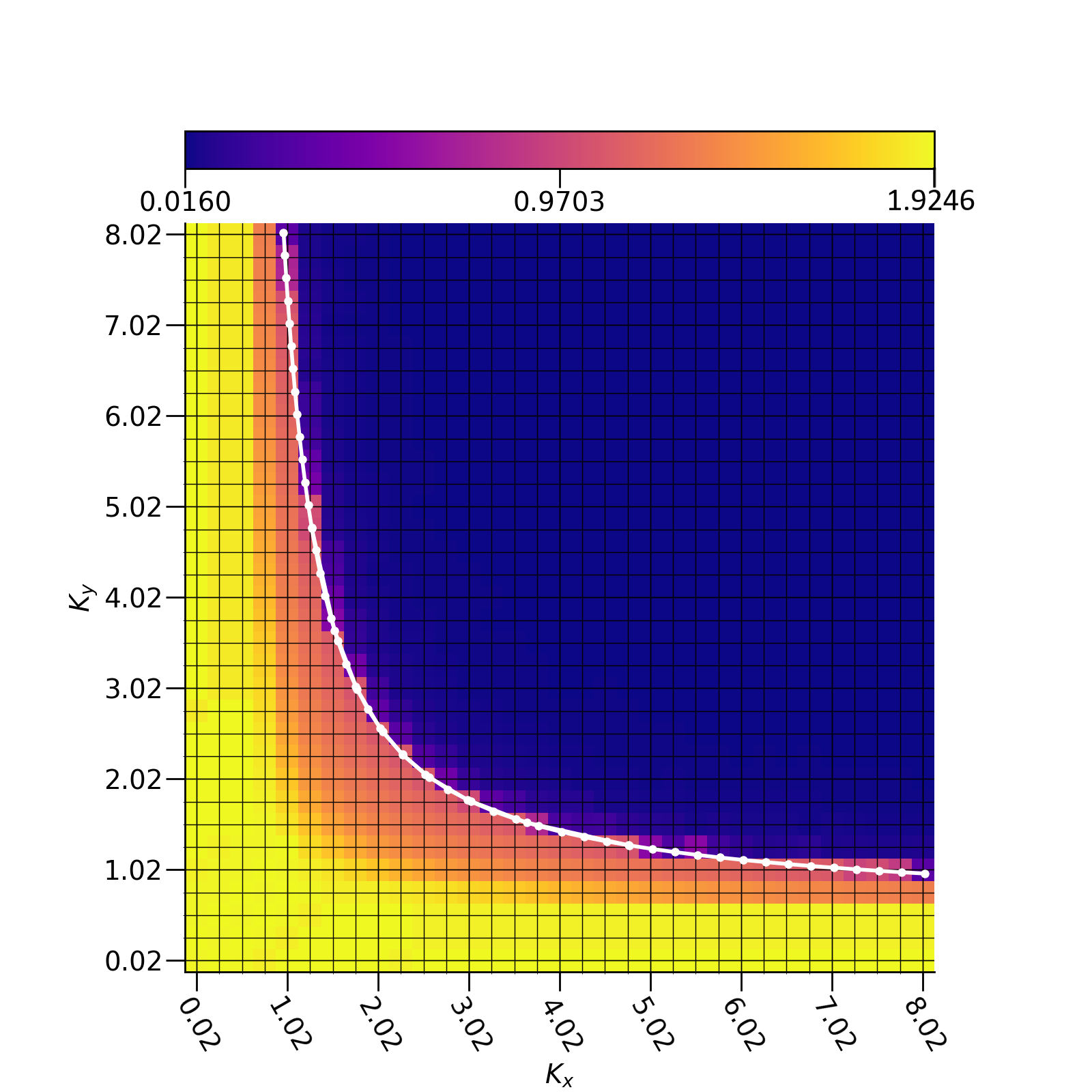} 
    \caption{The first principal component of the latent variance, $\tau_0$, with respect to the different $K_x$ and $K_y$ ($1/(\beta J_x)$ and $1/(\beta J_y)$) for the $2D$ square lattice. The left panel and right panel are for the system sizes $N=M=64$ and $N=M=128$, respectively.}
    \label{fig:4}    
    \end{figure*}

\begin{figure*}[tbp]
    \centering
    \includegraphics[width=8.525cm]{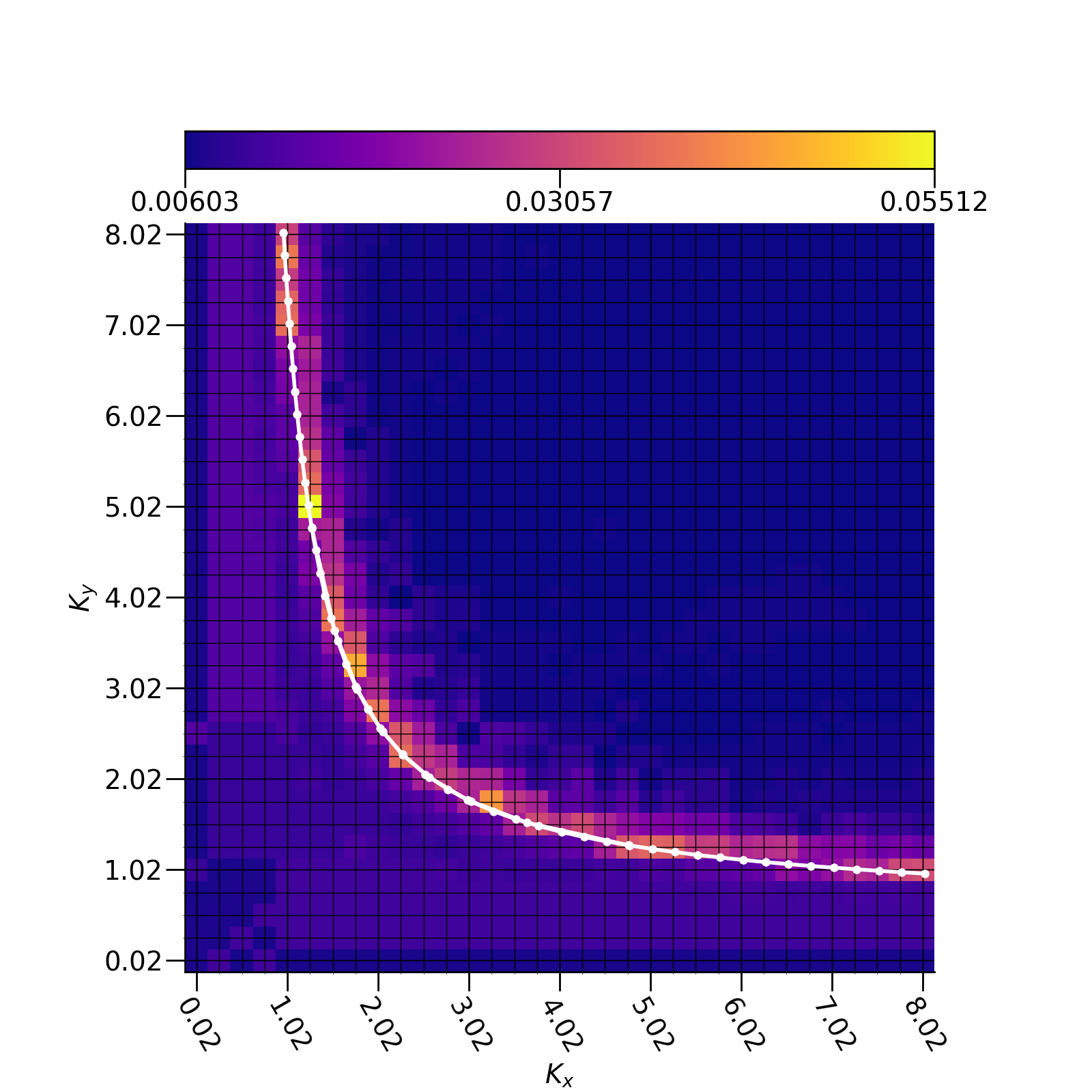} 
    \qquad
    \includegraphics[width=8.525cm]{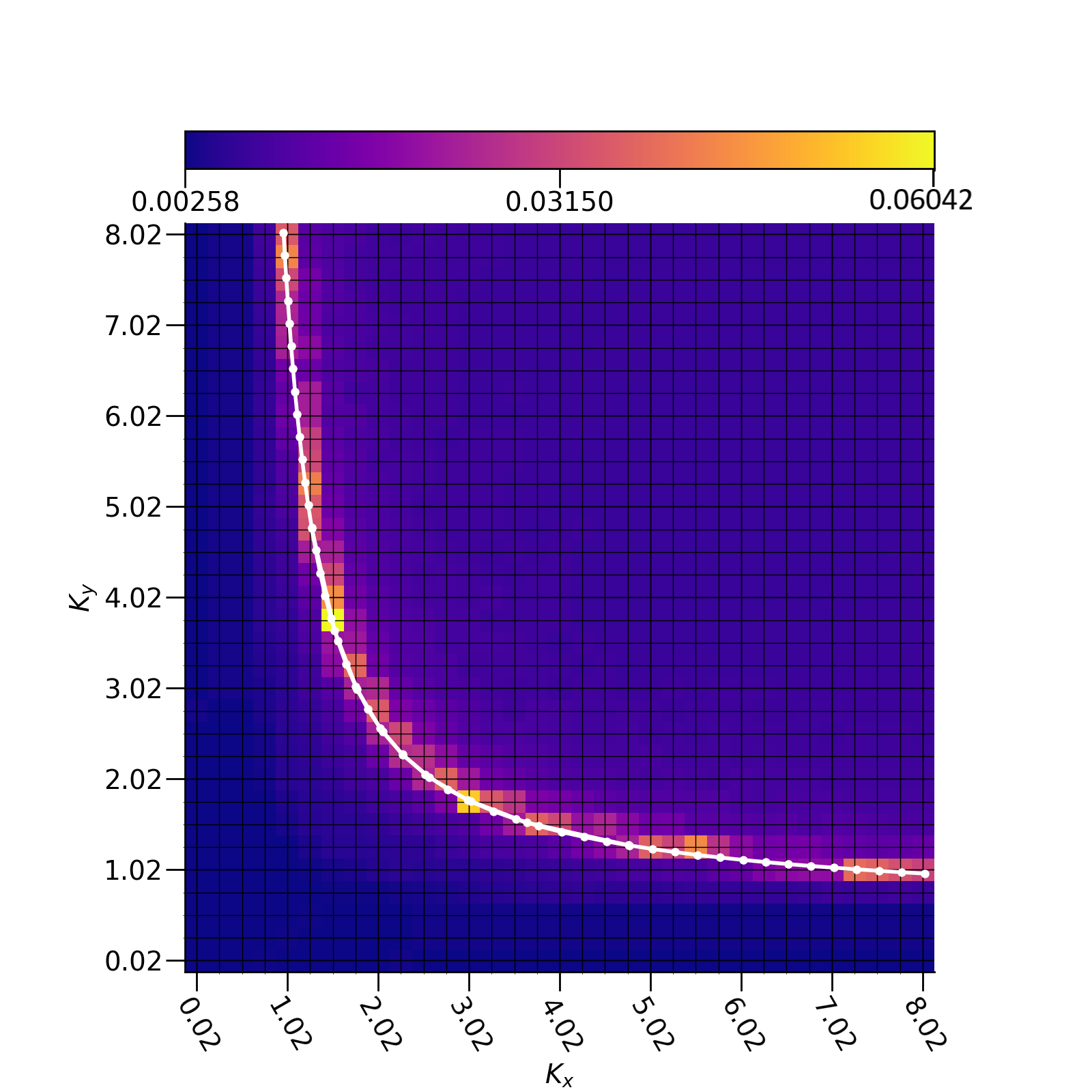} 
    \caption{The second principal component of the latent variance, $\tau_1$, with respect to the different $K_x$ and $K_y$ ($1/(\beta J_x)$ and $1/(\beta J_y)$) for the $2D$ square lattice. The left panel and right panel are for the system sizes $N=M=64$ and $N=M=128$, respectively. }
    \label{fig:6}
\end{figure*}

\begin{figure*}[tbh]
\centering
    \includegraphics[width=8.525cm]{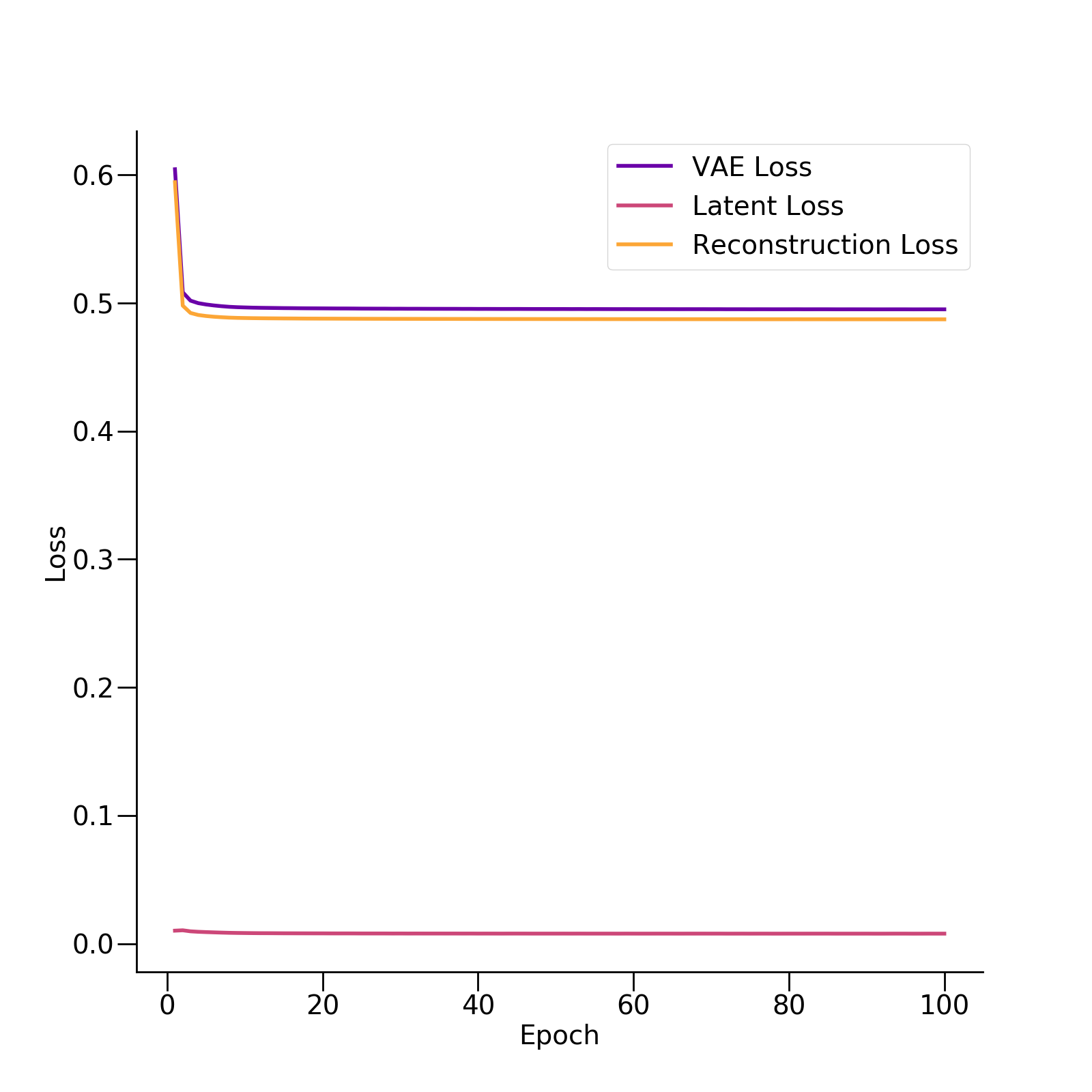} 
    \qquad
    \includegraphics[width=8.525cm]{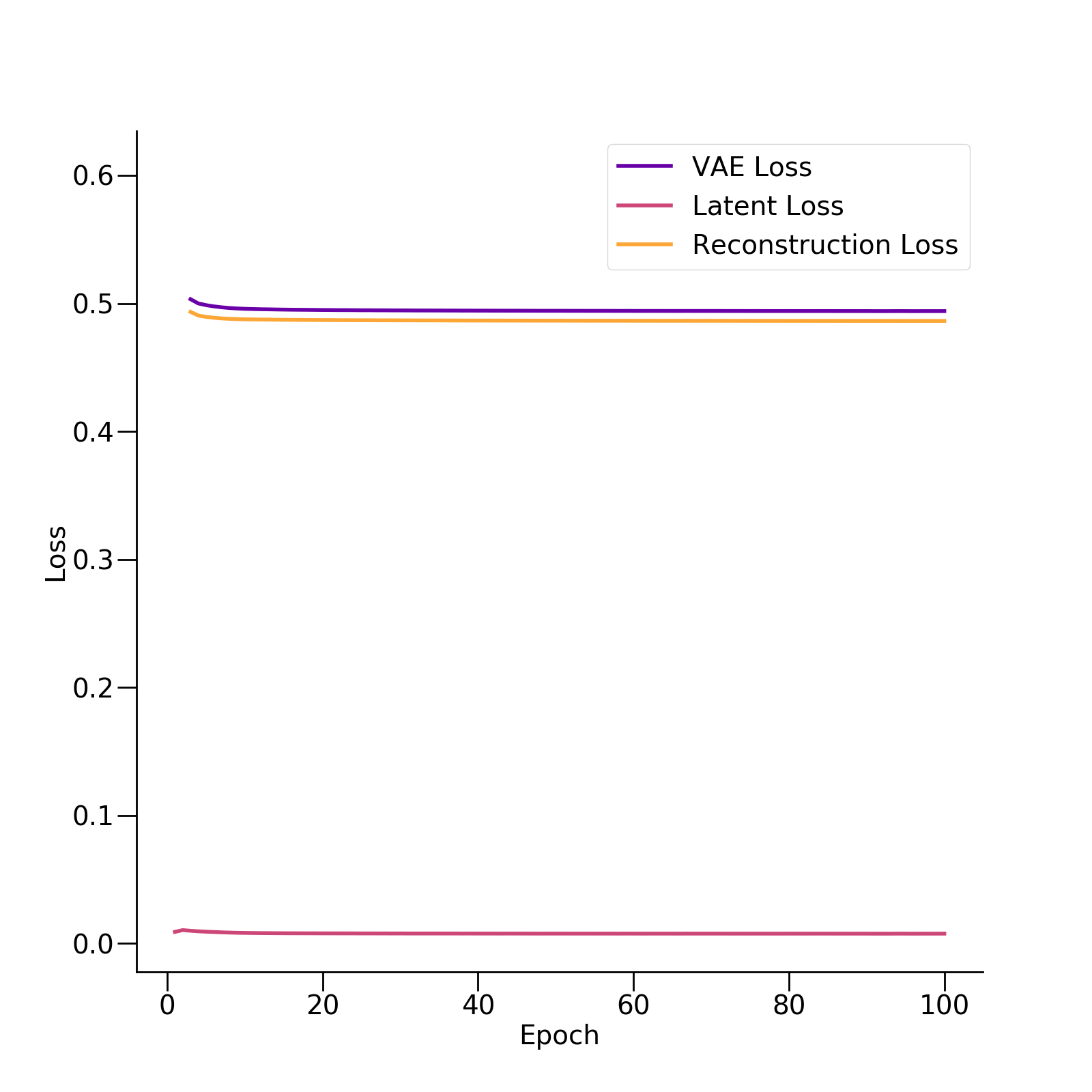} 
    \caption{$\beta$-TCVAE loss and the reconstruction loss settle to about $0.5$ and the latent loss settles close to $0$ for $100$ epoch cycles. The left panel and right panel are for the square lattice of size $N=M=64$ and $N=M=128$, respectively.}
\label{fig:8}
\end{figure*}

\end{center}

Given that the paramagnetic samples are essentially noise due to entropic contributions. Therefore, these happens to be easy to discriminate from the rest of the samples using a $\beta$-TCAVE model. In view of the fact, the samples with $\nu_1$ values corresponding to nearly zero magnetization and rather high values for $\tau_0$ resemble Gaussian noise with no notable order preference. An interesting question is to look for quantity which may resemble to response function, such as susceptibility and heat capacity. 

We plot the second principal component of the latent variance  $\tau_{1}$ in fig. \ref{fig:6}. The phase transition line is again included to show the transition in the thermodynamic limit for comparison. The values of $\tau_{1}$ are much larger along the analytical critical line than the other combinations of $K_{x}$ and $K_{y}$.
Thus $\tau_1$ bears a strong resemblance to the magnetic susceptibility.  

From the plots of $\tau_0$, $\tau_1$, and $\nu_1$, we find that the VAE is capable to generate quantities which resembles to the magnetization, amplitude of magnetization/energy, and magnetic susceptibility/heat capacity, respectively. From these quantities we can draw the phase transition line between the paramagnetic phase and the ferromagnetic phase. 

The quality of fitting the Ising spin configurations to the present VAE model can be quantified by the values of the loss functions.
The three losses represented in fig ~\ref{fig:8} are the VAE loss, reconstruction loss, and the latent loss. The reconstruction loss converges to about $0.5$. The latent loss is obtained from the $\beta$-TCKLD term. This loss converges quickly to a value near to $0$. The total $\beta$-TCVAE loss for the $2$-dimensional anisotropic Ising model for both lattice size $N=64$ and $N=128$ settles quickly around $0.5$.

\section{Discussion and Conclusions}

We use the $\beta$-TCVAE to extract structural information from raw Ising spin configurations. It exposes interesting derived descriptors of the configurations used to identify the second order phase transition line among other regions. The analysis here shows the interpretation of the extracted sample space as represented by the latent variables. This is done by studying the behavior of the latent variable mappings of the Ising spin configurations with respect to the anisotropic coupling $(J_x,J_y)$ associated with the temperatures. 

We find that $\nu_1$,  the second principal component of the latent mean, is considered to reflect the magnetization of the $2$-dimensional anisotropic Ising model, hence it is interpreted as an indicator for the magnetization exhibited by the configurations. In contrast, $\tau_0$ and $\tau_1$, which are the first two principal components of the latent variance, can be seen as an indicator of the amplitude of magnetization or energy and the magnetic susceptibility or heat capacity. Thus both $\tau_0$ and $\tau_1$ can provide a reasonable estimate of the second order phase transition line. 

Since the $d+1$-dimensional anisotropic Ising model is equivalent to the $d$-dimensional quantum spin system through the Suzuki-Trotter transformation. This method can be trivially extended to other 1D quantum systems \cite{10.1143/PTP.73.319}.

Moreover, methods for strongly correlated systems, such as the dynamical mean field theory (DMFT) and their cluster generalizations--the dynamical cluster approximation (DCA) and the cellular dynamical mean field theory (CDMFT)--have very similar data structure in their Hirsch-Fye impurity solver \cite{Maier_etal_2005,Georges_etal_1996,Hirsch_Fye_1986}. It is notoriously difficult to obtain the putative quantum critical point from the paramagnetic solution of the Hubbard model as there is no simple quantity to track the transition \cite{vidhyadhiraja2009quantum,kellar2020non}. Therefore, the method developed here would be an important tool for analyzing the data from the DMFT, DCA, and CDMFT.

There are many opportunities for further developing this method, not only investigating more complex systems, but also by introducing improvements which are beyond the scope of this work. Finite size scaling turns to be an important approach to address the limitations in solving finite-sized systems for investigating regions near critical phenomena \cite{fsc,Fisher_Barber_1972}. A correspondence established between the VAE encodings of different system sizes is a challenging argument, as different VAE structures need specific training for each system size, which in turn demands for different hyperparameters and training iterations, counts to obtain equivalent analysis and hence similar answers ~\cite{fsc}. Numerical difficulties tend to arise while performing finite-size scaling analysis, because the variation of predicted properties with respect to the system size is hard to isolate from the systemic variation due to different neural networks trained with different hyperparameters is being used to extract the specific macroscopic properties. However, this would play a significant role towards improving VAE characterization of critical phenomena.

Another interesting direction is to use the generative adversarial network (GAN)~\cite{Goodfellow_etal_2014} instead of VAE as the generative model, promising results have been obtained for the isotropic 2D Ising model \cite{Walker_Tam_2021}. 

\section{Acknowledgment}
This manuscript is based upon work supported by NSF DMR-1728457. This work used the high performance computational resources provided by the Louisiana Optical Network Initiative (http://www.loni.org), and HPC@LSU computing. NW research work at Louisiana State University was supported by NSF DMR-1728457. JM is partially supported by the U.S. Department of Energy, Office of Science, Office of Basic Energy Sciences under Award Number DE-SC0017861.

\appendix
\section{Self-Duality of the two-dimensional Ising Model}
In this appendix, we summarize the derivation of the critical points for the anisotropic $2$-dimensiaonl Ising model via the self-duality property of the model \cite{kwd_2,Kogut_1979,Muramatsu_2009}. We closely follow the lecture notes by Muramatsu in the following \cite{Muramatsu_2009}.

Considering the partition function of the Ising model (where $K=\frac{1}{\beta J}$)
\begin{multline}
    Z=\sum _{\{S_{j}\}}e^{K\sum_{<j,l>}S_{j}S_{l}}=\sum _{\{S_{j}\}}\prod_{<j,l>}e^{K S_{j}S_{l}} \\
    =\sum_{\{S_{j}\}}\prod_{<j,l>}\sum^{1}_{r=0}C_{r}(K)(S_{j}S_{l})^r
\end{multline}
where $C_{0}(K)=\cosh{K}$ and $C_{1}(K)=\sinh{K}$. Applying the simple transformation, for each bond $<j,l>$, a new $Z_2$ variable, $r$ is introduced. We label the new variable as $r_\mu$ with $\mu \equiv (i, <i, j>)$, labelling it with the site $i$ from which the bond $<i, j>$ emanates. The partition function thus follows
\begin{equation}
    Z=\sum_{\{S_j\}}\sum_{\{r_{\mu}\}}\prod_{<j,l>}C_{r_{\mu}}(K)\prod_{i}S^{\sum_{<i,j>}r_{\mu}}_{i}
\end{equation}
grouping all products of spins on site $i$ together, $\sum_{<i,j>}r_{\mu}$, contains all four contributions, resulted due to the bonds connected to site $i$. Further, we perform explicitly the sum over all spin configurations
\begin{multline}
    Z=\sum_{\{r_{\mu}\}}\prod_{<j,l>}C_{r_{\mu}}(K)\prod_{i}\sum_{S_{i}=\pm 1}S^{\sum_{<i,j>}r_{\mu}}_{i} \\
    =\sum_{\{r_{\mu}\}}\prod_{<j,l>}C_{r_{\mu}}(K)\prod_{i}2\delta[\mod_{2}(\sum_{<i,j>}r_{\mu})]
\end{multline}
We define a dual lattice, where the vertices of the dual lattice are set in the center of the plaquettes defined by the original lattice. We have vanishing contributions due to the presence of the Kronecker delta in many configurations. Defining the new $Z_2$ variables $\sigma_i = \pm1$ on the sites of the dual lattice. We can associate for each link of the original lattice, there is a pair of $\sigma_{i}$’s (e.g. on the sites $i$ and $j$ on the dual lattice). Therefore, the variable $r_{\mu}$ is expressed as:
\begin{equation}
    r_{\mu}=\frac{1}{2}(1-\sigma_{i} \sigma_{j})
\end{equation}
where sites $i$ and $j$ on the dual lattice are those where the link crosses $r_{\mu}$. The sum of $r_{\mu}$ is over the four nearest neighbors of a site $i$, we have 
\begin{equation}
    \sum_{<i,j>}r_{\mu}=2-\frac{1}{2}(\sigma_{1}\sigma_{2}+\sigma_{2}\sigma_{3}+\sigma_{3}\sigma_{4}+\sigma_{4}\sigma_{1})
\end{equation}
There are $2^4$ possible configurations for the four variables $\sigma_{1},...,\sigma_{4}$. They are grouped in four cases and all the cases lead to an even number for the sum of $r_{\mu}$ over the nearest neighbor bonds. The choice of variables needs to satisfy the $\delta-$function. The partition function becomes
\begin{equation}
    Z= \frac{1}{2}2^{N}\sum_{\{\sigma_{i}\}}\prod_{<j,l>}C_{[(1-\sigma_{i}\sigma_{j})/2]}(K),
\end{equation}
where $N$ is the number of sites on the lattice and the product is now over the bonds associated in the dual lattice. The expression of the partition function shows that the weight for each configuration of the $\sigma$'s is given by the coefficients $C(K)$. Hence, introducing $C(K)$ to a form that resembles a Boltzmann weight. 
\begin{center}
\begin{eqnarray}
    C_{r}(K)&&= \cosh(K)[1+r(\tanh(K)-1)] \\ \nonumber 
    &&=\cosh(K) \exp(\ln[1+r(\tanh(K)-1)]) \\ \nonumber
    &&=\cosh(K) \exp(r \ln \tanh(K))\\ \nonumber
    &&=\cosh(K) \exp[\frac{1}{2}(1-\sigma_{i}\sigma_{j}) \ln \tanh(K)] \\ \nonumber 
    &&=[\cosh(K) \sinh(K)]^{1/2} \exp(-\frac{1}{2} \ln \tanh K \sigma_{i} \sigma_{j})
\end{eqnarray}
\end{center}
The partition function becomes
\begin{equation}
    Z=1/2(\sinh 2 \Tilde{K})^{-N}\sum_{\{\sigma_{i}\}}exp(\Tilde{K}\sum_{<j,l>}\sigma_{j}\sigma_{l}).
\end{equation}
There are $2N$ bonds and defining the new coupling constant as
\begin{equation}
   \Tilde{K} \equiv -\ln \tanh(K/2)  
\end{equation}
(where $K=\frac{1}{\beta J}$) coupling for the Ising model on the dual lattice.  The Ising model is self-dual since the duality transformations brings it into itself. We consider the free energy per site 
\begin{equation}
    f(Z)=-\frac{1}{N} \ln Z.
\end{equation}
According to the relation between the partition functions of the original and the dual model, we can write
\begin{equation}
    f(K)=\ln \sinh (2 \Tilde{K}) + f(\Tilde{K}).
\end{equation}
This is a strong constraint on the free energy. Since, $\sinh(2\tilde{K})$ is an analytic function, the equation has a singularity in $f(K)$, which corresponds to a
singularity in $f(\tilde{K})$. $\tilde{K}(K)$ is a monotonous function of $K$, hence it holds $\tilde{K_c}=K_c$. We have
\begin{equation}
    K_c=\frac{1}{2}\ln(1+\sqrt{2}).
\end{equation}
The self-duality has allowed us to calculate the exact value of the critical temperature in the $2$-dimensional isotropic Ising model$(K_x=K_y)$, where $K_x=\frac{1}{\beta J_x}$ and $K_y=\frac{1}{\beta J_y}$. Generalizing the results obtained in the isotropic to the anisotropic one, i.e. when couplings $K_x \neq K_y$ in the respective directions. The anisotropic case is as follows
\begin{equation}
    \tilde{K_{y}} \equiv -\frac{1}{2}\ln \tanh(K_x)  ,  \tilde{K_x} \equiv -\frac{1}{2}\ln \tanh(K_y).
\end{equation}
Given $K_x$ and $K_y$ there is only one critical point, summarizing both the equations above with the following condition for a critical line separating the ordered from the disordered phase in the anisotropic Ising model
\begin{equation}
    \sinh(2K_{xc})\sinh(2K_{yc})=1\label{eq:exact}.
\end{equation}

\section{$\beta$-TCVAE Loss}
The expression of the total loss for the VAE is given by
\begin{equation}
    L(\phi,\theta ;x,z)=L_{RC}+L_{KLD},
\end{equation}
where the reconstruction error (RC) and the Kullback-Leibler divergence (KLD) are defined as 
\begin{equation}
   L_{RC}=-E_{z \sim q_\phi(z|x)}\left[ \log p_\theta(x|z) \right]
\end{equation}
and 
\begin{equation}
   L_{KLD}=D_{KL}\left[q_\phi(z|x)||p(z)\right]
\end{equation}
respectively. If the prior distribution of the latent representation is Gaussian, the VAE provides disentangled factors in the latent representation, which means the significant dimension of the latent space is largely independent of each other. In $\beta$-total correlation VAE ($\beta$-TCVAE), we try to improve the disentanglement of factors in the representation by decomposing the KLD term and apply tuning parameters to them independently ~\cite{chen2019isolating,walker2020identifying}. Each  training sample is identified with a unique integer index $n \in 1,2,..,N $ and assigned a uniform random variable in this decomposition. The aggregated posterior, $q_\phi(z)=\sum_n q_\phi(z|n) p(n)$ captures the aggregate structure of the latent variables under the distribution of the input, where $q_\phi(z|n)=q_\phi(z|x_n)$ and $q_\phi(z,n)=q_\phi(z|n)p(n)=\frac{1}{N}q_\phi(z|n)$. The decomposition is given as
\begin{eqnarray}
  I(z;x)+D_{KL}[q_\phi(z)||\prod_{j}q_\phi(z_{j})] \nonumber \\ +\sum_{j}D_{KL}[q_\phi(z_{j})||p(z_{j})]
\end{eqnarray}
The first term is the index-code mutual information (MI), $I(z;x)=D_{KL}[q_\phi(z,n)||q_\phi(z)p(n)]$, between the input and the latent variable, which is based on the empirical input distribution $q_\phi(z,n)$. The second term is a measure of the dependence between the latent variables and is called the total correlation(TC). It is important to produce representations which penalize the total correlation and hence force the model towards discovering statistically disentangled factors in the input distribution. The third term prevents the individual latent variables in the representation from deviating far from their priors. It is called the dimension-wise KLD ~\cite{walker2020identifying}. After adding the tuning parameters to the decomposition, the $\beta$-TC modified KLD term becomes :- 
\begin{eqnarray}
   L_{\beta-TC}= \alpha I(z;x) 
  + \beta D_{KL}[q_\phi(z)||\prod_{j}q_\phi(z_{j})] \nonumber \\
  + \gamma \sum_{j}D_{KL}[q_\phi(z_{j})||p(z_{j})]
\end{eqnarray}
Modulating only the $\beta$ parameter shows the greatest effect on disentanglement in the latent representation given by empirical evidence ~\cite{chen2019isolating}.

\medskip

\bibliography{references.bib}

\newpage

\end{document}